\documentclass[aps,prb,reprint,amsmath,amssymb,superscriptaddress,floatfix]{revtex4-2}

\usepackage{graphicx}
\usepackage{dcolumn}
\usepackage{bm}
\usepackage{multirow}
\usepackage{subfigure}
\usepackage{booktabs}
\usepackage{hhline}  
\usepackage{float}
\usepackage{braket}
\usepackage[colorlinks,linkcolor=blue,anchorcolor=blue,citecolor=blue,urlcolor=blue]{hyperref}

\begin{document}
\title{
  Non-perturbative \textit{ab initio} approach for calculating the electrical conductivity of a liquid metal}

\author{Xiao-Wei Zhang}
\thanks{These two authors contribute equally.}
\affiliation{International Center for Quantum Materials, Peking University,
  Beijing 100871, P. R. China}

\author{Haoran Chen}
\thanks{These two authors contribute equally.}
\affiliation{International Center for Quantum Materials, Peking University,
  Beijing 100871, P. R. China}

\author{En-Ge Wang}
\affiliation{International Center for Quantum Materials, Peking University,
  Beijing 100871, P. R. China}
\affiliation{Ceramic Division, Songshan Lake Lab, Institute of Physics, Chinese Academy of Sciences, Guangdong, China}
\affiliation{School of Physics, Liaoning University, Shenyang, China}

\author{Junren Shi}
\email{junrenshi@pku.edu.cn}
\affiliation{International Center for Quantum Materials, Peking University,
  Beijing 100871, P. R. China}
\affiliation{Collaborative Innovation Center of Quantum Matter, Beijing 100871, P. R. China}

\author{Xin-Zheng Li}
\email{xzli@pku.edu.cn}
\affiliation{Interdisciplinary Institute of Light-Element Quantum Materials and Research Center for Light-Element Advanced Materials, State Key Laboratory for Artificial Microstructure and Mesoscopic Physics, Frontier Science Center for Nano-optoelectronics and School of Physics, Peking University, Beijing 100871, China}
\affiliation{Peking University Yangtze Delta Institute of Optoelectronics, Nantong, Jiangsu 226010, China}
\date{\today}

\begin{abstract}
We propose a non-perturbative \textit{ab initio} approach to calculate the electrical conductivity of a liquid metal. 
Our approach is based on the Kubo formula and the theory of electron-phonon coupling (EPC), and unlike the conventional empirical approach based on the Kubo-Greenwood formula, fully takes into account the effect of coupling between electrons and moving ions. 
We show that the electrical conductivity at high temperature is determined by an EPC parameter $\lambda_{\mathrm{tr}}$, which can be inferred, non-perturbatively, from the correlation of electron scattering matrices induced by ions. 
The latter can be evaluated in a molecular dynamics simulation. 
Based on the density-functional theory and pseudopotential methods, we implement the approach in
an \textit{ab initio} manner. 
We apply it to liquid sodium and obtain results in good agreement with experiments. 
This approach is efficient and based on a rigorous theory, suitable for applying to general metallic liquid systems.
\end{abstract}
\maketitle

\section{Introduction}
Liquid metals are an important class of materials vital for many applications because of their excellent electrical and thermal conductivities combined with flexible mechanic properties.
They find applications in flexible electronics~\cite{mohammed_all-printed_2017,bo_recent_2018,eaker_oxidation-mediated_2017}, microfluidic devices~\cite{dickey_emerging_2014,eaker_liquid_2016} and material syntheses~\cite{daeneke_liquid_2018}, etc.
They also form cores of many planets, generating geomagnetic fields~\cite{desjarlais_electrical_2002,holst_electronic_2011,pozzo_thermal_2012}.
The physical properties of liquid metals are crucial for these applications and for understanding the formations and evolution of planets.
Among them, electrical conductivity is a basic nonetheless one of the most important properties.
As such, an efficient and reliable approach for calculating the electric conductivity of a liquid metal is highly desirable.

At present, the most successful approach for calculating the electrical conductivity of a liquid metal 
is to combine molecular dynamics (MD) or path-integral molecular dynamics (PIMD) simulations with the Kubo-Greenwood (KG) formula~\cite{kubo_statistical-mechanical_1957, greenwood_boltzmann_1958}.
This approach approximates liquid as an ensemble of independent electron systems subject to quenched and disordered ionic fields, and inherently ignores the effect of the motion of ions on the evolution of electron states~\cite{dufty_kubo-greenwood_2018}.
The electrical conductivity is calculated by averaging ion configurations sampled from MD simulations~\cite{allen_electrical_1987,silvestrelli_electrical-conductivity_1997,desjarlais_electrical_2002,dufty_kubo-greenwood_2018}.
The direct-current (dc) limit ($\omega\rightarrow0$) of the conductivity is obtained by extrapolating 
from conductivities at high-frequencies~\cite{silvestrelli_electrical-conductivity_1997,desjarlais_electrical_2002, pozzo_electrical_2011}. 
This approach has been applied to various liquid metals with great successes~\cite{silvestrelli_electrical-conductivity_1997,knider_ab_2007,pozzo_electrical_2011,pozzo_thermal_2012,pozzo_transport_2013,koker_electrical_2012,vlcek_electrical_2012,pozzo_saturation_2016,drchal_transport_2017,korell_paramagnetic--diamagnetic_2019}.
However, its approximated and empirical nature limits further improvements. 
Parallel to this, in the more rigorous treatment developed for solids, electrical resistivity can be
interpreted as a result of electron-phonon coupling (EPC)~\cite{pozzo_saturation_2016,drchal_transport_2017,korell_paramagnetic--diamagnetic_2019}, 
for which the dynamic effects of ion motion play a central role. 
The harmonic approximation of ion motion, however, is often applied and the EPC is treated in a 
perturbative manner.
Both of these treatments are not applicable in liquids.
It is not obvious how the two distinct views can be unified in a certain limit. 
Nor it is satisfactory that one has to rely on two distinct theories for two phases of one matter.
In this paper, we extend the applicability of the EPC theory of conductivity originally developed for solids and develop a non-perturbative \textit{ab initio} approach appropriate for calculating the dc conductivity of a liquid. 
Instead of perturbatively determining the EPC scattering amplitudes, we reformulate the theory to relate the dc conductivity with the irreducible interaction mediated by EPC in the particle-hole channel.
We show that the irreducible interaction can be inferred from the correlation of the $\mathcal{T}$ matrices of electron-ion scatterings, which can be evaluated in a MD or PIMD simulation.
At high temperatures, the formula of electrical resistivity is reduced to the familiar form of the conventional EPC theory that is proportional to a single EPC parameter $\lambda_{\mathrm{tr}}$. 
We explicitly relate the parameter to the irreducible interaction. 
With these developments, we have a non-perturbative approach for determining electrical conductivity.
We implement the approach by using the density-functional theory (DFT) and norm-conserving pseudopotential methods. 
Applying the implementation to liquid sodium, we find that the resistivity of sodium has an upward jump when transited from a solid to a liquid phase, and exhibits non-linear temperature-dependence at high temperatures. 
The results are in good agreement with available experiments, both qualitatively and quantitatively.
The remainder of the paper is organized as follows. 
In Sec.~\ref{Gen_theory}, we develop the general formalism of the electrical conductivity in liquids. 
The relation between the conductivity and the irreducible interaction ${I}$ mediated by EPC in the electron-hole channel is established. 
In Sec.~\ref{Interaction}, we show how ${I}$ can be inferred from a MD simulation by relating it to the correlation of $\mathcal{T}$ matrices of electron-ion scatterings.
The approach is applied to liquid sodium. 
Implementation details and results are presented in Sec.~\ref{Application}. 
Finally, we summarize and discuss our results in Sec.~\ref{Conclusion}.
Some details of the theoretical derivations and the tests of numerical convergence are presented in Appendices.

\section{EPC theory of electrical conductivity}\label{Gen_theory}
In this section, we develop formalism for calculating the dc conductivity of a liquid. The derivation is based on the conventional EPC theory of conductivity originally developed for solids, see, e.g., Ref.~\onlinecite{mahan_many-particle_2013}.  To have a theory appropriate for a liquid, we need to eliminate reliance on perturbatively defined quantities and the harmonic approximation in the original theory.
\subsection{Formula of conductivity}
From the Kubo formula~\cite{kubo_statistical-mechanical_1957,mahan_many-particle_2013}, the dc electrical conductivity of a general system can be calculated by
\begin{eqnarray}
  \sigma=-\lim_{\omega\rightarrow0}\frac{\mathrm{Im}[\pi_\mathrm{ret}(\omega)]}{\omega},
\end{eqnarray}
where $\pi_\mathrm{ret}(\omega)$ is the retarded current-current correlation function. To determine $\pi_\mathrm{ret}(\omega)$, it is more convenient to first determine the imaginary-time-ordered correlation function
\begin{eqnarray}
  \pi(i\omega_m)=-\frac{1}{3V}\int_0^{\hbar\beta}d\tau e^{i\omega_m\tau}\left\langle\hat{T}_\tau\hat{\bm{j}}(\tau)\cdot\hat{\bm{j}}(0)\right\rangle,
\end{eqnarray}
and then perform an analytic continuation by substituting $i\omega_m$ with $\omega+i\delta$, where $\delta$ denotes an infinitesimal positive constant. Here $\tau\in[0,\hbar\beta)$ is the imaginary time, with $\beta=1/k_B T$ being the inverse temperature. $\omega_m=2m\pi/\hbar\beta, m\in Z$ is a Boson Matsubara frequency. $V$ is the total volume of the system. We have assumed that the system is isotropic.

By substituting the current operator $
  \hat{\bm{j}}_{\bm q}(\tau)=-({e}/{m})\sum_{\bm{p}\sigma}\bm{p}
  \hat{\psi}_{\bm{p}\sigma}^\dagger(\tau)\hat{\psi}_{\bm{p}\sigma}(\tau)
$ into the Kubo formula, we obtain:
\begin{equation}\label{pi_ori}
  \pi({i}\omega_m)
  =\frac{2e^2}{3m^2V}\frac{1}{\hbar\beta}\sum_{pp'}
      \bm{p}\cdot\bm{p}'{G}^{(2)}(p,p'+q_0;p+q_0,p'),
\end{equation}
where
\begin{multline}
{G}^{(2)}(p,p'+q_0;p+q_0,p')
  =-\int_{0}^{\hbar\beta}d\tau e^{i\omega_{m}\tau}\\
   \times\left\langle\hat{T}_{\tau}
      \hat{\psi}_{\bm{p}\sigma}^\dagger(\tau + \tau_{0})
      \hat{\psi}_{\bm{p}\sigma}(\tau + \tau_{0})
      \hat{\psi}_{\bm{p'}\sigma}^\dagger(\tau_{0})
      \hat{\psi}_{\bm{p'}\sigma}(\tau_{0})\right\rangle
\end{multline}
is the two-particle Green's function, $\hat{\psi}_{\bm{p}\sigma}$ ($\hat{\psi}_{\bm{p}\sigma}^\dagger$) is the annihilation (creation) operator of an electron with the momentum $\bm{p}$ and spin $\sigma$, $e$ and $m$ are the charge and bare mass of the electron, respectively. For simplicity, we use the 4-dimensional momentum notation $p\equiv(\bm{p},i\nu)$, and $q_0\equiv(\bm{0},i\omega_m)$, where $\nu\equiv (2n+1)\pi/\hbar\beta,n\in Z$ denotes a Fermion Matsubara frequency.

The two-particle Green's function ${G}^{(2)}$ can in general be decomposed as
\begin{multline}
  \label{eq:G2Gamma}
  {G}^{(2)}(p,p'+q_0;p+q_0,p') =
  {{G}}_{p}{{G}}_{p+q_0}\delta_{pp'}\\
  +\frac{1}{\hbar^{2}\beta} {{G}}_{p}{{G}}_{p+q_0}
  \Gamma_{pp'}(q_{0}) {{G}}_{p'}
  {{G}}_{p'+q_0},
\end{multline}
where we ignore the inconsequential disconnected part of the Green's function,  ${{G}}_{p}$ is the single-particle Green's function of the system, and  $\Gamma_{pp'}(q_0)$ is the scattering amplitude of an electron-hole pair scattered from $(p,p+q_{{}_0})$ to $(p',p'+q_0)$. We note that ${{G}}_{p}$ is diagonal in the basis of plane waves because a liquid has the space and time translation symmetries.

\begin{figure}[tbp]
\centerline{\includegraphics[width=3.4in]{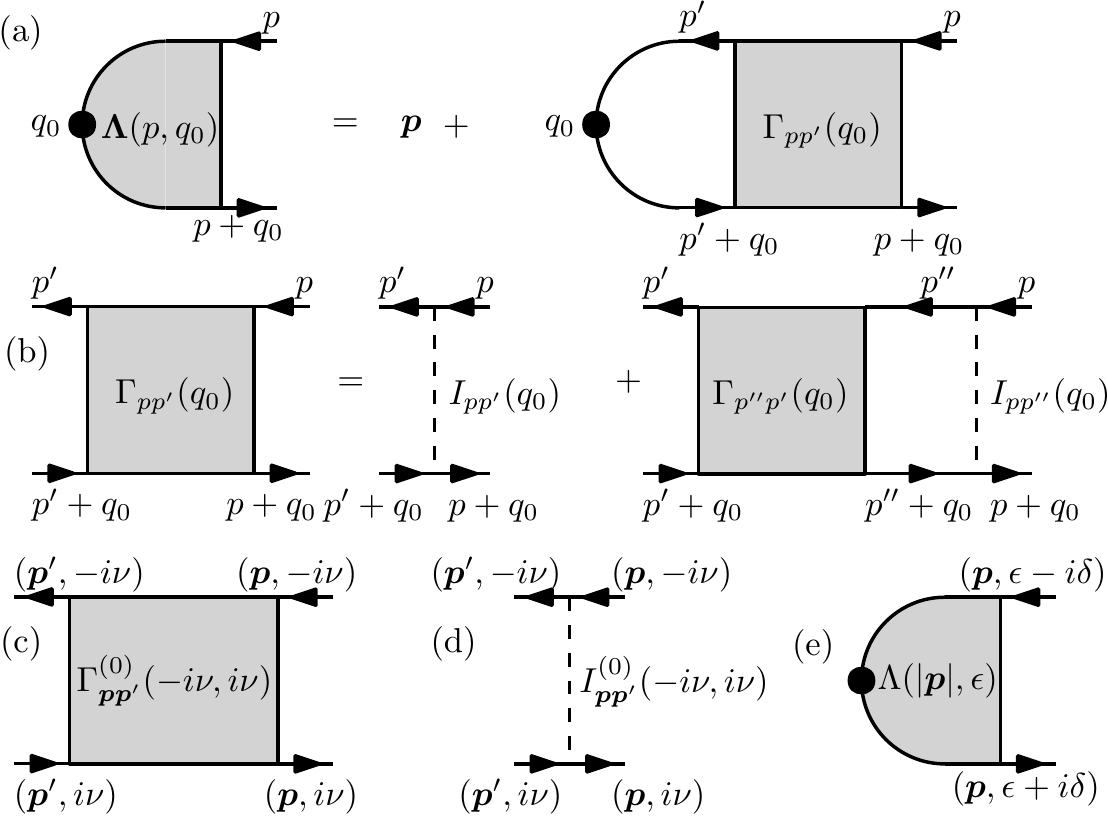}}
\caption[]{\label{fig:feyn} (a) The Bethe-Salpeter equation (\ref{BSE}). (b) The relation between the vector vertex function and the scattering amplitude. (c-e) The definitions of $\Gamma^{(0)}_{\bm p \bm p'}(-i\nu,i\nu)$, $I^{(0)}_{\bm p \bm p'}(-i\nu, i\nu)$, and $\Lambda(|\bm p|, \epsilon)$, respectively.}
\end{figure}

We define a vector vertex function $\bm\Lambda(p, q_{0}) = \bm{p}+ ({1}/{\hbar^2\beta}) \sum_{p'} \bm{p}' {{G}}_{p'} {{G}}_{p'+q_0} \Gamma_{pp'}(q_0)$ [see Fig.~\ref{fig:feyn}(a)].  Because a liquid is isotropic, the vector vortex function must have the form $\bm\Lambda(p, q_{0}) = \bm p \Lambda(|\bm p|; i\nu, i\nu+i\omega_{m})$, where $\Lambda$ is a scalar vertex function~\cite{mahan_many-particle_2013}.  Equation~(\ref{pi_ori}) can be rewritten as
\begin{equation}\label{pi_sgamma}
  \pi(i\omega_m)=\frac{2e^2}{3m^2}\frac{1}{\hbar\beta}\sum_{p}
  {{G}}_{p}{{G}}_{p+q_0}
  |\bm{p}|^2\Lambda(|\bm{p}|;i\nu,i\nu+i\omega_m).
\end{equation}

We then complete the summation of the Matsubara frequency $\nu$, and preform the analytic continuation $i\omega_{m} \rightarrow \omega + i\delta$.  After applying Ward identities for EPC systems~\cite{engelsberg1963,mahan_many-particle_2013}, we obtain
\begin{multline}
\label{eq:sigma1}
\sigma = \frac{2e^{2}}{3m^{2}} \int \frac{d^{3}p}{(2\pi\hbar)^{3}}\left|\bm p\right|^{2}\int_{-\infty}^{\infty} \frac{d\epsilon}{2\pi\hbar}  \left[- \frac{d n_{F}(\epsilon)}{d\epsilon}\right]\\
\times \left|G^{\mathrm{ret}}_{\bm p}(\epsilon)\right|^{2} \Lambda(|\bm p|, \epsilon - i\delta, \epsilon + i\delta),
\end{multline}
where ${G}^{\mathrm{ret}}_{\bm p}(\epsilon)$ is the retarded Green's function. The details of the analytic continuation can be found in \S8.4.2 of Ref.~\onlinecite{mahan_many-particle_2013}.

We further apply the approximation
\begin{eqnarray} \label{eq:Gnorm}
  \left|{G}_{\bm p}^{\mathrm{ret}}(\epsilon) \right|^{2}
  \approx -\frac{\pi\hbar^2}{\mathrm{Im}{\Sigma}(|\bm p|, \epsilon+i\delta)} \delta(\epsilon+\epsilon_F-\tilde{\epsilon}_{\bm p}(\epsilon)),\quad
\end{eqnarray}
where $\Sigma(|\bm p|, \epsilon+i\delta)$ is the self-energy of the system, $\tilde{\epsilon}_{\bm{p}}(\epsilon)=(|\bm p|^2/2m)+\mathrm{Re}{\Sigma}(|\bm p|,\epsilon)$ is the renormalized electron dispersion, and $\epsilon_F$ is the Fermi energy. The approximation is valid when $|\mathrm{Im}\Sigma(|\bm p|, \epsilon+i\delta)| \ll \epsilon_{F}$, which is true for most metallic systems.

Finally, noting that the system is isotropic and the vertex function and the self-energy only weakly depend on $|\bm p|$ for $|\bm p|\sim p_F$, we obtain the formula for determining the dc conductivity of a liquid
\begin{eqnarray}\label{conduct}
  \sigma=\frac{e^2n_0\hbar}{2m}\int_{-\infty}^{\infty}d\epsilon \frac{\Lambda(p_F;\epsilon-i\delta,\epsilon+i\delta)}{Z(\epsilon)\mathrm{Im}{\Sigma}(p_F;\epsilon+i\delta)}\frac{d n_F(\epsilon)}{d\epsilon},\quad\quad
\end{eqnarray}
where $n_0$ is the electron density, $n_F(\epsilon)$ is the Fermi-Dirac distribution function, and $p_{F}$ is the Fermi momentum. $\Lambda(p_F;\epsilon-i\delta,\epsilon-i\delta)$ is obtained from $\Lambda(p_F;i\nu,i\nu+i\omega_m)$ by substituting $i\nu\rightarrow\epsilon-i\delta$ and $i\nu+i\omega_m\rightarrow\epsilon+i\delta$, and $Z(\epsilon) = 1+ ({m}/{p_{F}}) [{\partial \mathrm{Re}\Sigma(p, \epsilon+i\delta)}/{\partial p}]_{p=p_{F}}$ is a factor due to the renormalization of the electron dispersion.

\subsection{Integral equations}
While Eq.~(\ref{conduct}) has a form identical to that of the conventional EPC theory, all complexities are hidden in the scalar vertex function.  In the conventional theory, the vertex function is determined perturbatively from an EPC Hamiltonian based on the harmonic approximation.  For liquids, instead, we make use of exact integral relations.

For the scattering amplitude $\Gamma$, we have the Bethe-Salpeter equation [Fig.~\ref{fig:feyn}(b)]~\cite{negele_quantum_1988}
\begin{multline}\label{BSE}
  \Gamma_{p,p+q}(q_0)
  =I_{p,p+q}(q_0)
  +\frac{1}{\hbar^2\beta}\sum_{q'}
  I_{p,p+q'}(q_0)\\
  \times{{G}}_{p+q'}
  {{G}}_{p+q'+q_0}
  \Gamma_{p+q',p+q}(q_0),
\end{multline}
where we introduce an irreducible electron-hole interaction $I$. In the perturbation theory, $I$ includes all the two-particle scattering diagrams that are irreducible in the direct electron-hole channel~\cite{giuliani_quantum_2005}. For liquids, we have to determine it non-perturbatively. This will be discussed in the next section.

From the relation between the vertex functions and $\Gamma$ [Fig.~\ref{fig:feyn}(a)], it is straightforward to obtain the integral equation for the scalar vertex function
\begin{multline}\label{sgamma}
  \Lambda(|\bm{p}|;i\nu,i\nu+i\omega_m)=1+
  \frac{1}{\hbar^2\beta}\sum_{q}
  \frac{\bm{p}\cdot\left(\bm p+\bm q\right)}{|\bm{p}|^2}
  I_{p,p+q}(q_0)\\
  \times
  {{G}}_{p+q}{{G}}_{p+q+q_0}
  \Lambda(|\bm p+ \bm q|;i\nu,i\nu+i\omega_m).
\end{multline}

\subsection{High temperature limit}\label{highT_1}
In most cases, the melting temperature of a material is much higher than its Debye temperature $\Theta_{\mathrm{D}}$. As a result, it suffices to determine the conductivity at the high-temperature limit $T \gg \Theta_{\mathrm{D}}$.  In this case, the conductivity is determined by a single EPC parameter $\lambda_{\mathrm{tr}}$.  This is shown as follows.

In the high temperature limit, Eq.~(\ref{sgamma}) can be simplified. Note that the Matsubara frequency $\hbar\omega_{m_{q}}\equiv2m_{q}\pi k_B T\gg k_{B}\Theta_{\mathrm{D}}$ unless $m_{q}=0$, while $I_{p,p+q}(q_{0})$, which is induced by the EPC, has significant magnitude only when $\hbar\omega_{m_{q}} \lesssim k_{B}\Theta_{\mathrm{D}}$, where $\omega_{m_{q}}$ denotes the Matsubara frequency of $q$. We therefore keep only terms with $\omega_{m_{q}}=0$ in the summation in Eq.~(\ref{sgamma}).
 As a result, for $i\nu\rightarrow\epsilon-i\delta$ and $i\nu+i\omega_m\rightarrow\epsilon+i\delta$, the equation can be simplified as
\begin{multline}\label{sgamma_pe}
  \Lambda(|\bm{p}|;\epsilon)=1+
  \frac{1}{\hbar^2\beta}\sum_{\bm{q}}
  \frac{\bm{p}\cdot\left(\bm p+\bm q\right)}{|\bm p|^2}
  I_{\bm{p},\bm p+\bm q}^{(0)}(\epsilon-i\delta,\epsilon+i\delta)\\
  \times \left|{G}_{\bm p+\bm q}^{\mathrm{ret}}(\epsilon)\right|^{2} \Lambda(|\bm p+\bm q|;\epsilon),
\end{multline}
where $\Lambda(|\bm{p}|;\epsilon)\equiv\Lambda(|\bm{p}|;\epsilon-i\delta,\epsilon+i\delta)$ [Fig.~\ref{fig:feyn}(e))], and $I_{\bm{p},\bm p + \bm q}^{(0)}(\epsilon-i\delta,\epsilon+i\delta)$ denotes $I_{p,p+q}(q_{0})$ for the given set of the momenta and a zero frequency transfer [Fig.~\ref{fig:feyn}(d))].

We then apply the approximation Eq.~(\ref{eq:Gnorm}). It gives rise to a Dirac delta function which constrains $\bm p + \bm q$ on the Fermi surface, i.e.,  $|\bm p + \bm q|= p_F$.  As a result, $\Lambda(|\bm p + \bm q|; \epsilon) = \Lambda(p_{F}; \epsilon)$ can be moved out of the summation. We define a set of EPC parameters
\begin{multline}\label{lambda}
  \begin{pmatrix} \lambda(\epsilon) \\ \lambda_{\mathrm{tr}}(\epsilon) \end{pmatrix} = \sum_{\bm{q}}
  \begin{pmatrix} 1 \\ -\bm p \cdot \bm q / |\bm p|^{2} \end{pmatrix}
  I_{\bm{p},\bm p+\bm q}^{(0)}(\epsilon-i\delta,\epsilon+i\delta)\\
  \times Z(\epsilon) \delta(\epsilon+\epsilon_F-\tilde{\epsilon}_{\bm p+\bm q}(\epsilon)).
\end{multline}
By using the parameters, the imaginary part of the self energy can be written as
\begin{eqnarray}\label{ImS}
  \mathrm{Im}{\Sigma}(\bm{p},\epsilon)
  \approx-\frac{\pi}{\beta Z(\epsilon)}\lambda(\epsilon)
\end{eqnarray}
in the high temperature limit (see Appendix~\ref{ImSE}). It is then straightforward to get the solution
\begin{eqnarray}\label{gamma_final}
  \Lambda(p_F;\epsilon)
  =\frac{\lambda(\epsilon)}{\lambda_{\mathrm{tr}}(\epsilon)}.
\end{eqnarray}

By inserting Eq.~(\ref{ImS}) and Eq.~(\ref{gamma_final}) into Eq.~(\ref{conduct}), we determine the dc conductivity. At high temperature and for $\epsilon\sim0$, we can neglect the energy dependence of $\lambda_{\mathrm{tr}}(\epsilon)$~\cite{takegahara1977}. Completing the integral over $\epsilon$, we obtain
\begin{eqnarray}\label{sigma_highT_s}
  \sigma&&\approx\frac{e^2n_0\hbar\beta}{2\pi m \lambda_{\mathrm{tr}}(0)}.
\end{eqnarray}
This is the final formula to be applied for determining the dc conductivity of a liquid. It has a form identical to that of the conventional theory. However, for liquids, the EPC parameter $\lambda_{\mathrm{tr}}(0)$ cannot be determined in a perturbative way.  According to Eq.~(\ref{lambda}), to determine $\lambda_{\mathrm{tr}}(0)$, we need to first determine the irreducible electron-hole interaction $I_{\bm{p},\bm p + \bm q}^{(0)}(-i\delta,i\delta)$.


\section{Irreducible electron-hole  interaction}\label{Interaction}
From the last section, we see that the irreducible electron-hole interaction $I_{\bm{p},\bm p + \bm q}^{(0)}(-i\delta,i\delta)$ is the key for determining the electrical conductivity of a liquid. In this section, we develop an approach for determining it.

\subsection{Related to an  electron-hole scattering amplitude}
By setting $p=(\bm p, i\nu)$, $p+q_{0}=(\bm p, -i\nu)$ and $q=(\bm q, 0)$ in Eq.~(\ref{BSE}), we obtain an equation for $I_{\bm{p},\bm p + \bm q}^{(0)}(-i\nu,i\nu)$:
\begin{multline}\label{BSE_static}
  I_{\bm{p},\bm p + \bm q}^{(0)}(-i\nu,i\nu)
  =\Gamma_{\bm{p},\bm p + \bm q}^{(0)}(-i\nu,i\nu)\\
  -\frac{1}{\hbar^2\beta}\sum_{\bm{q'}}
  I_{\bm{p},\bm p + \bm q'}^{(0)}(-i\nu,i\nu)\\
  \times \left|{{G}}_{\bm p + \bm q'}(-i\nu)\right|^2
  \Gamma_{\bm p + \bm q',\bm p + \bm q}^{(0)}(-i\nu,i\nu),
\end{multline}
where $\Gamma_{\bm{p},\bm p + \bm q}^{(0)}(-i\nu,i\nu)$ denotes $\Gamma_{p,p+q}(q_{0})$ for the given set of the momenta [Fig.~\ref{fig:feyn}(c)], and we keep only terms with zero Matsubara frequency in the summation of the right hand side, as it is appropriate for the high-temperature limit.

The equation suggests an approach for determining the irreducible interaction: by determining $\Gamma_{\bm{p},\bm p + \bm q}^{(0)}(-i\nu,i\nu)$ numerically, we can obtain $I_{\bm{p},\bm p + \bm q}^{(0)}(-i\nu,i\nu)$ by solving Eq.~(\ref{BSE_static}). It is reasonable to expect that $I_{\bm{p},\bm p + \bm q}^{(0)}(-i\nu,i\nu)$ only weakly depends on the Matsubara frequency $i\nu$, the irreducible interaction can be obtained by
\begin{equation}
  I_{\bm{p},\bm p + \bm q}^{(0)}(-i\delta,i\delta) \approx I_{\bm{p},\bm p + \bm q}^{(0)}(-i\nu,i\nu),
\end{equation}
for a properly chosen $i\nu$ (see below).

\subsection{Evaluating the scattering amplitude in MD}\label{sec:eval-scatt-ampl}
Quantities like $\Gamma_{\bm{p},\bm p + \bm q}^{(0)}(-i\nu,i\nu)$, which is defined in the imaginary time, can in general be evaluated in a PIMD simulation. In the simulation, one maps quantum ion degrees of freedom into classical ring polymers with beads representing ions at different instances of the imaginary time~\cite{chandler_exploiting_1981}. Electron-related quantities can be evaluated by averaging an ensemble of quantum electron systems subjected to random imaginary-time-dependent ionic fields. Applications of such an approach can be found in Ref.~\onlinecite{liu_superconducting_2020,chen_first-principles_2021}.

%
In the approach, the single-particle can be evaluated as $G_{p}=\langle\mathcal{G}_{pp}[\bm{R}(\tau)]\rangle_{C}$, and the two-particle Green's function as
\begin{multline}\label{G2_PIMD}
  {G}^{(2)}(p+q_{0},p';p,p'+q_{0})\\
  =\left\langle\mathcal{G}_{pp'}[\bm{R}(\tau)]
    \mathcal{G}_{p'+q_0,p+q_{0}}[\bm{R}(\tau)]
  \right\rangle_C,
\end{multline}
where $\mathcal{G}[\bm{R}(\tau)]$ denotes the electron Green's function at a given ion configuration $\{\bm{R}_i(\tau)\}$, and  $\langle\cdots\rangle_C$ denotes an average over ion configurations.

The scattering amplitude $\Gamma_{pp'}(q_{0})$ can be expressed as a correlation function. To see that, we apply the identity
\begin{eqnarray}\label{Tmatrix}
  \mathcal{G}_{pp'}={{G}}_{p}\delta_{p,p'}
  +\frac{1}{\hbar}{{G}}_{p}\mathcal{T}_{pp'}{{G}}_{p'},
\end{eqnarray}
where $\mathcal{T}_{pp'}$ denotes the matrix element of the $\mathcal{T}$ matrix of electron scattering induced by an ionic field. Substitute Eq.~(\ref{Tmatrix}) into Eq.~(\ref{G2_PIMD}), and compare the resulting form with Eq.~(\ref{eq:G2Gamma}), we find:
\begin{equation}\label{Gamma}
  \Gamma_{pp'}(q_{0})=\beta\left\langle
  \mathcal{T}_{pp'}
  \mathcal{T}_{p+q_{0},p'+q_{0}}
  \right\rangle_C.
\end{equation}

We can show that the $\mathcal{T}$-matrix has the symmetry
\begin{equation}
  \label{Tsymmetry}
\left(\mathcal{T}_{p,p+q}\right)^{\ast} = \mathcal{T}_{\bar{p}+q,\bar{p}},
\end{equation}
where we denote $\bar{p}\equiv(\bm p, -i\nu)$. To see this, we note that the Green's function can be determined by the matrix equation $[\mathcal{G}^{-1}]_{pp'} = {G}_{0}^{-1}(p)\delta_{pp'} - V_{\bm{pp'}}(i\nu-i\nu')$, where ${G}_{0}(p)$ is the Green's function in a free space, and $V_{\bm{pp'}}(i\nu-i\nu')$ is the Fourier transform of the random ionic potential. Since the ionic potential is Hermitian and is a local function of the time, we have $V_{\bm{pp'}}^{\ast}(i\nu-i\nu') = V_{\bm{p'p}}(i\nu'-i\nu)$. In addition, we have $[{G}_{0}(p)]^{\ast}={G}_{0}(\bar{p})$. By applying these relations, it is straightforward to show $\left(\mathcal{G}_{p,p+q}\right)^{\ast} = \mathcal{G}_{\bar{p}+q,\bar{p}}$. Besides, time reversal and inversion symmetries require that $[{{G}}(p)]^{\ast}={{G}}(\bar{p})$. Combining the relations with Eq.~(\ref{Tmatrix}), we obtain Eq.~(\ref{Tsymmetry}).

Applying Eq.~(\ref{Gamma}) and Eq.~(\ref{Tsymmetry}), we have
\begin{eqnarray}\label{Gamma_simp}
  \Gamma^{(0)}_{\bm p, \bm p + \bm q}(-i\nu, i\nu)=\beta\left\langle
  |\mathcal{T}_{\bar{p},\bar{p}+q}|^2
  \right\rangle_C
\end{eqnarray}
with $q\equiv(\bm q, 0)$. This is the formula to be applied for evaluating the scattering amplitude.

For a PIMD simulation, to determine the $\mathcal{T}$-matrix for a time-dependent ionic potential, one needs to solve the time-dependent equation of the Green's function
\begin{equation}\label{GreenG}
  \left[- \frac{\partial}{\partial\tau}-\frac{\hat{H}(\tau)-\epsilon_F\hat{\mathbb{I}}}{\hbar}\right] \hat{\mathcal{G}}(\tau,\tau^\prime) = \delta(\tau-\tau^\prime )\hat{\mathbb{I}},
\end{equation}
where $\hat{H}(\tau)$ denotes the time-dependent Hamiltonian for a given ionic potential.  This is expensive and infeasible in practice.  Fortunately, we can apply the quasi-static approximation~\cite{liu_superconducting_2020}. This is to choose a Matsubara frequency $\nu$ with its magnitude $k_{B}\Theta_{\mathrm{D}}/\hbar\ll |\nu| \ll \epsilon_{F}/\hbar$, and determine instantaneous solutions $\hat{\mathcal{G}}(i\nu;\tau) = [(i\nu + \epsilon_{F}/\hbar)\hat{\mathbb{I}} - \hat H(\tau)/\hbar]^{-1}$.  The approximated solution of Eq.~(\ref{GreenG}) can then be written as
\begin{equation}
\label{eq:quasi-static}
\hat{\mathcal{G}}(i\nu+i\omega_{m}, i\nu) \approx \frac{1}{\hbar\beta}
\int d\tau \hat{\mathcal{G}}(i\nu;\tau)e^{i\omega_{m}\tau},
\end{equation}
where $\hat{\mathcal{G}}(i\nu+i\omega_{m}, i\nu)$ denotes the Fourier transform of $\hat{\mathcal{G}}(\tau,\tau')$.  The $\mathcal{T}$-matrix can be obtained by applying Eq.~(\ref{Tmatrix}).

Finally, since the scattering amplitude $\Gamma^{(0)}_{\bm p, \bm p + \bm q}(-i\nu, i\nu)$ has a zero frequency transfer, and in most cases, ions are heavy enough to have a negligible effect of quantum fluctuations, it is usually sufficient to use the classical MD instead of the more expensive PIMD for simulating the motion of ions. In this case, the $\mathcal{T}$-matrix can be obtained straightforwardly from the static ionic potential with respect to a given ionic configuration.

\section{Application to liquid sodium}\label{Application}
In this section, we implement and apply our approach to liquid sodium.

\subsection{Implementation}
In our implementation, \textit{ab initio} MD and PIMD simulations are performed using the Quantum Espresso package interfaced with i-PI~\cite{giannozzi_quantum_2009,ceriotti_i-pi_2014}. The Martins-Troullier norm-conserving pseudopotential is used to treat the ion-electron interactions~\cite{troullier_efficient_1991}. The Perdew-Burke-Ernzerhof (PBE) functional is used to describe the exchange-correlation potential~\cite{perdew_generalized_1996}. The electron Brillouin zone is sampled with the $\Gamma$-point. An energy cutoff of 30 Ry is used for the expansion of electron wave functions by plane waves. At 400 K, the MD simulations are run for supercells containing 128, 250 and 432 atoms, and the PIMD simulations are run for a supercell of 250 atoms with 4 and 8 beads. From 500 to 800 K, MD simulations are run for the supercell of 250 atoms. The time step is 3 fs and the simulation time is not less than 24 ps. The Generalized Langevin equation (GLE) thermostat is used to equilibrate the canonical ensemble. For each simulation, after the temperature reaches equilibrium, the nuclear configurations are uniformly sampled with a spacing of 25 time steps. To get a converged result, we usually need $\sim 400$ samples. The atomic densities at different temperatures were set according to the experiment~\cite{osminin_density_1965}.

With ionic configurations output by the MD or PIMD simulations, we determine the Green's function $\mathcal{G}[\bm{R}(\tau)]$ for each of the configurations. Corresponding $\mathcal{T}$ matrices are determined by applying Eq.~(\ref{Tmatrix}). By averaging the ionic configurations, we obtain the single-particle Green's function and the scattering amplitude $\Gamma^{(0)}_{\bm p, \bm p + \bm q}(-i\nu, i\nu)$. The irreducible interaction $I^{(0)}_{\bm p, \bm p + \bm q}(-i\nu, i\nu)$ is obtained from the scattering amplitude by solving Eq.~(\ref{BSE_static}).

To determine $\lambda_{\mathrm{tr}}(0)$, we recast $I^{(0)}_{\bm p, \bm p + \bm q}(-i\nu, i\nu)$ for $\bm p$ and $\bm p + \bm q$ close to the Fermi surface as a function of $q\equiv|\bm q|$. By interpolation, we can have a function $I^{(0)}(q)$ for arbitrary values of $q$.  The EPC parameters can then be determined by
\begin{eqnarray}
  \begin{pmatrix} \lambda(0) \\ \lambda_{\mathrm{tr}}(0)\end{pmatrix} =
  {N(\epsilon_F)} \int_{0}^{2p_F}
  \begin{pmatrix}{q}/{2 p_{F}^2} \\[0.5em] {q^{3}}/{4 p_{F}^4}\end{pmatrix}
  I^{(0)}(q)dq,
    \label{lambda_int}
\end{eqnarray}
where $N(\epsilon_{F})$ is the density of states of free electrons at the Fermi Surface.


\subsection{Results}

\begin{figure}[htb]
  \includegraphics[width=\columnwidth]{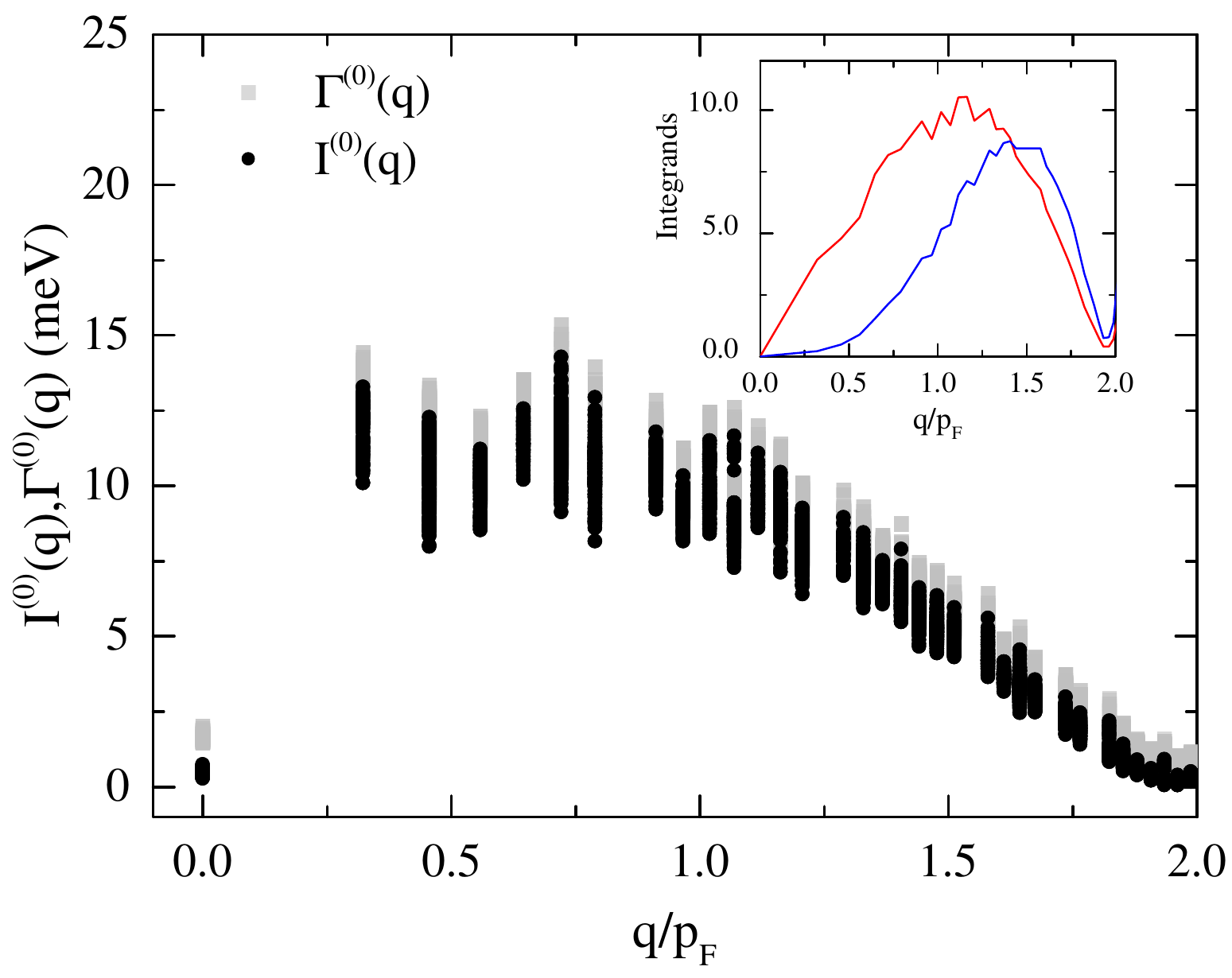}
  \caption{\label{fig-WGamma}The irreducible electron-electron interaction $I^{(0)}(q)$ and the scattering amplitude $\Gamma^{(0)}(q)$ of liquid sodium  at 400~K.  The functions are recast from $I^{(0)}_{\bm p\bm p'}$ and $\Gamma^{(0)}_{\bm p \bm p'}$ for  $q=|\bm{p}-\bm{p'}|$ and  $0.9p_{F} < |\bm{p}|, |\bm p'| < 1.1p_{F}$, respectively. Values for different $\bm p$'s and $\bm p'$'s but having the same $q$ are shown as separated points. The vertical spreads of the values indicate their uncertainties. The inset shows the integrands of Eq.~(\ref{lambda_int}) for $\lambda_{\mathrm{tr}}(\lambda)$ . }
\end{figure}

In Fig.~\ref{fig-WGamma}, we show the irreducible electron-hole interaction $I^{(0)}_{\bm p, \bm p + \bm q}(-i\nu, i\nu)$ recast as a function $I^{(0)}(q)$ for liquid sodium at 400 K. The result shown is determined from a 36-ps MD simulation for a supercell of 250 atoms, where 460 samples are extracted. PIMD simulations with different number of beads are also performed, and yield no statistically distinguishable changes (see Appendix~\ref{sec:tests-convergence}). Due to the finite size of the simulation, there is no data for $0<q\lesssim 0.3k_{F}$, although we do find $I^{(0)}(0)\approx 0$ as expected for an interaction induced by EPC.  The lack of data in the region introduces uncertainty for interpolating values of $I^{(0)}(q)$.  Fortunately, the uncertainty will not severely affect the determination of $\lambda_{\mathrm{tr}}(0)$, as the contribution from the region is suppressed by the $q^{3}$ factor in Eq.~(\ref{lambda_int}) (see the inset of Fig.~\ref{fig-WGamma}). We also show the scattering amplitude $\Gamma^{(0)}(q)$ from which $I^{(0)}(q)$ is inferred. It is numerically close to $I^{(0)}(q)$ but with a notable difference: $\Gamma^{(0)}(0)\ne 0$. It suggests that applying the Bethe-Salpeter equation (\ref{BSE_static}) is important for recovering the correct asymptotic behavior of the irreducible interaction.

By applying Eq.~(\ref{lambda_int}), we calculate the EPC parameters $\lambda(0)$ and $\lambda_{\text{tr}}(0)$ of liquid sodium at 400 K. The result is shown in Table~\ref{tab:result}.
%
%
%
Values for solid sodium, both from the conventional EPC theory~\cite{bauer_electron-phonon_1998} and experiments~\cite{elliott_haas-van_1982,allen_empirical_1987}, are also shown.
%
%
We find that liquid sodium has a value of $\lambda_{\mathrm{tr}}(0)$ nearly twice as large as that of solid sodium. It suggests a large enhancement of EPC when sodium transits from the solid to the liquid phase. The similar enhancements of EPC parameters are also found in amorphous solids~\cite{grimvall_electron-phonon_1981,bergmann_eliashberg_1971}, which share similar static structure as liquids.  According to Eq.~(\ref{sigma_highT_s}), the enhancement of $\lambda_{\mathrm{tr}}(0)$ will induce a jump of the resistivity in the solid-liquid transition.

\begin{table}[tbp]
  \begin{tabular*}{1\columnwidth}{@{\extracolsep{\fill}} cccc}
    \toprule
    & \multicolumn{1}{c}{\text{Liquid}} & \multicolumn{2}{c}{Solid (bcc)}  \\
    \midrule
    \midrule
    \multirow{3}{*} & This work & $\text{Theory}^{\mathrm{a}}$ & \text{Experiment} \\
    $\lambda(0)$ & $0.36(3)$ & $0.18$ & $0.218^{\mathrm{b}}$ \\
    $\lambda_{\mathrm{tr}}(0)$ & $0.23(1)$ & $0.12$ & $0.14^{\mathrm{c}}$ \\
    \bottomrule
  \end{tabular*}
  \caption{\label{tab:result}%
    The EPC parameters $\lambda(0)$ and $\lambda_{\text{tr}}(0)$ of liquid sodium at 400 K in comparison with previous theoretical and experimental results of solid sodium.  The numbers in the parenthesis denote estimated uncertainties. The superscripts ``a'', ``b'', ``c'' denote data from Ref.~\onlinecite{bauer_electron-phonon_1998}, Ref.~\onlinecite{elliott_haas-van_1982}, and Ref.~\onlinecite{allen_empirical_1987}, respectively. }
\end{table}

\begin{figure}[hbt]
  \includegraphics[width=\columnwidth]{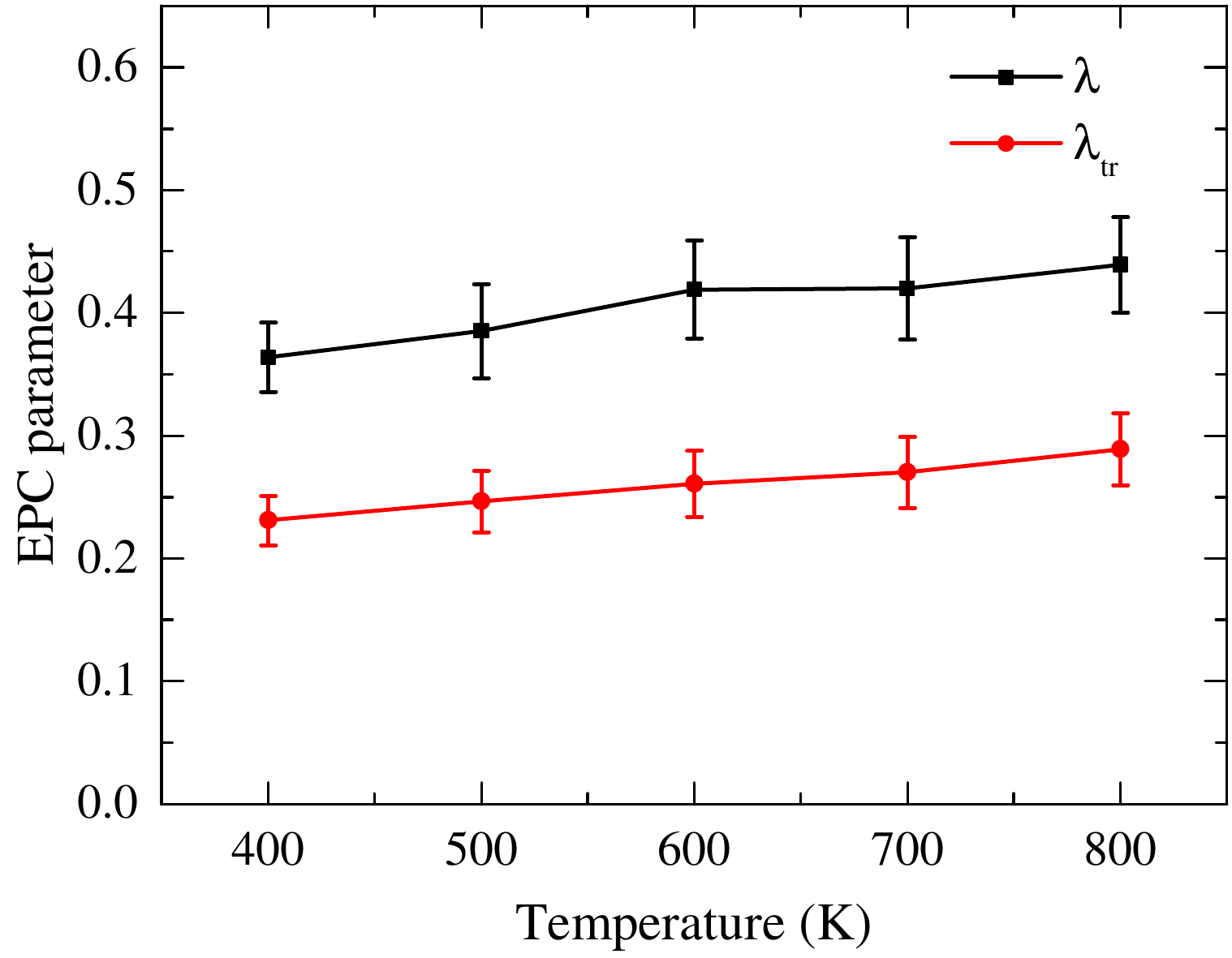}
  \caption{\label{fig-lambda}The temperature dependence of EPC parameters $\lambda(0)$ and $\lambda_{\text{tr}}(0)$ of liquid sodium. The errors are estimated from the vertical spreads of the values of $I(q)$ shown in Fig.~\ref{fig-WGamma}.
  }
\end{figure}

%
We calculate the EPC parameters at different temperatures and show the results in Fig.~\ref{fig-lambda}.  It is evident that the EPC parameters in liquid sodium are temperature-dependent.  In the conventional EPC theory based on the harmonic approximation, these parameters are temperature independent.  As a result, the resistivity of solid shows linear temperature dependence at high temperatures.  In contrast, the resistivity of liquid sodium will show non-linear temperature dependence because of the temperature dependence of $\lambda_{\mathrm{tr}}(0)$.

\begin{figure}[tbp]
  \includegraphics[width=\columnwidth]{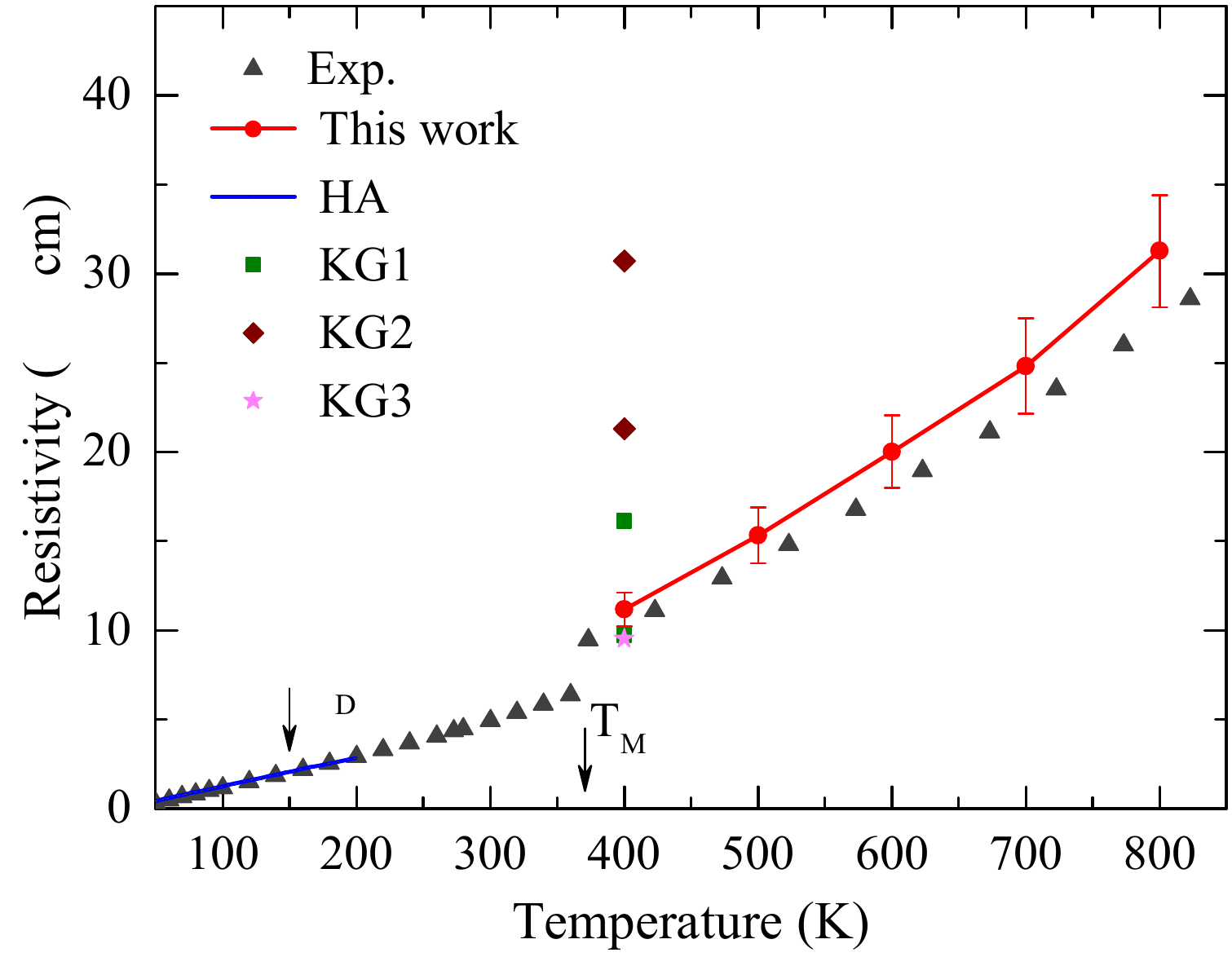}
  \caption{\label{fig-rho} Temperature dependence of the resistivity of sodium. We show the theoretical results of this work, previous theoretical results from the Kubo-Greenwood method (KG1--3, from Ref.~\onlinecite{pozzo_electrical_2011}, Ref.~\onlinecite{silvestrelli_electrical-conductivity_1997}, and Ref.~\onlinecite{knider_ab_2007}, respectively), the linear temperature dependence predicted by the conventional EPC theory based on the harmonic approximation (HA) (the value of $\lambda_{\mathrm{tr}}$ is from Ref.~\onlinecite{bauer_electron-phonon_1998}), as well as experimental data. 
  The resistivity variations among previous works at 400 K are caused by the choices of DFT exchange-correlation functional, k-point sampling, and supercell size.
  The experimental data for the liquid and the solid phases are from Ref.~\onlinecite{freedman1961} and Ref.~\onlinecite{bass1983electrical}, respectively.
  }
\end{figure}

Finally, we show the temperature-dependence of the electrical resistivity of liquid sodium and compare our theoretical results with experimental measurements in Fig.~\ref{fig-rho}.  The agreement is good in both the magnitude and the trend.  The theory correctly predicts the upward jump of the resistivity at the melting point with a magnitude coinciding well with the experimental observation.  The theory also correctly predicts the non-linear dependence of the resistivity observed in experiments.  The quantitative differences between the theory and the experiments are within the error bars of the current calculation.

%
%
%
%

\section{Summary and discussion}\label{Conclusion}
In conclusion, we have developed a non-perturbative approach to calculate the electronic resistivity of a liquid.  We show that the resistivity is determined by a single EPC parameter $\lambda_{\mathrm{tr}}$ at high temperature. We further show that the EPC parameter can be related to the irreducible electron-hole interaction ${I}$, which can be inferred from the fluctuation of scattering $\mathcal{T}$-matrices induced by the coupling to ions. The fluctuation of the $\mathcal{T}$-matrices can be determined from a MD simulation. To verify the approach, we develop an \textit{ab initio} implementation based on DFT and pseudopotential methods and apply it to liquid sodium. The theoretical results are in good agreement with experiments.

%
Compared to the conventional approach based on the Kubo-Greenwood formula~\cite{kubo_statistical-mechanical_1957,greenwood_boltzmann_1958}, our new approach has a number of advantages. Firstly, our approach is more efficient. The conventional approach determines the dc conductivity by extrapolation from conductivities at finite frequencies. In low frequencies, the approach requires a large supercell for obtaining a converging result. For example,  previous simulations have to employ a supercell containing as large as 2,000 atoms for liquid sodium~\cite{pozzo_electrical_2011}. In contrast, our approach calculates the dc conductivity directly from the irreducible interaction which is expected to be short-range, and the finite size effect is not as severe. Actually, a 250-atom supercell already gives a satisfactory result in our calculation. Secondly, our approach is based on rigorous formalism instead of an empirical method. The calculation of the dc electric conductivity of a liquid shares a unified theoretical ground with the same calculation for its solid phase. Finally, our approach can be improved. We can identify approximations involved in our approach such as taking the high-temperature limit and ignoring the $\epsilon$ dependence of the vertex function. These approximations can be scrutinized and improved if necessary.

%
%
%
%
%
%
%
%
%
%
%

\begin{acknowledgments}
  The authors are supported by the National Basic Research Programs of China under Grand Nos. 2016YFA0300900, 2017YFA0205003 and 2018YFA0305603, the National Science Foundation of China under Grant Nos 11774003, 11934003, 11888101, 11634001 and 12174005, the Strategic Priority Research Program of Chinese Academy of Sciences under Grant No. XDB28000000, and Beijing Municipal Science \& Technology Commission under Grant No. Z181100004218006. The computational resources were provided by the High-performance Computing Platform of Peking University.
\end{acknowledgments}


\appendix

\section{Imaginary part of the self energy}\label{ImSE}
%

In the theory of liquid superconductivity~\cite{liu_superconducting_2020}, Liu et al. introduce an effective electron-electron interaction $W$.  Comparing the equations satisfied by $I^{(0)}(-i\nu, i\nu)$ (see Sec.~\ref{sec:eval-scatt-ampl}) and those for $W$~\cite{liu_superconducting_2020}, we conclude
\begin{equation}\label{eq:IW}
  I^{(0)}_{\bm p,\bm p'}(-i\nu,i\nu)=-{W}_{11^{\prime}}
\end{equation}
with $1\equiv (\bm p, i\nu)$ and $1^{\prime} = (\bm p', i\nu)$.

Liu et al. also establish a generalized optical theorem for the imaginary part of the self-energy
\begin{eqnarray}\label{eq:ot}
  \mathrm{Im}{\Sigma}_1=-\frac{1}{\hbar\beta}\sum_{1'} (\mathrm{Im}{{G}}_{1'}) W_{1'1}.
\end{eqnarray}
The relation is exact.

We can apply these two relations to determine the imaginary part of the self-energy at high temperature. In this case, we have
\begin{equation}
  \mathrm{Im}\bar{\Sigma}(\bm{p},i\nu)
  \approx \frac{1}{\hbar\beta}\sum_{\bm{q}} \mathrm{Im}{G}(\bm{p}',i\nu) I_{\bm p'\bm{p}}^{(0)}(-i\nu,i\nu).
\end{equation}
Applying the analytic continuation $i\nu\rightarrow \epsilon+i\delta$ and the approximation $\mathrm{Im}G(\bm p', \epsilon+i\delta) \approx -\pi\hbar\delta(\epsilon + \epsilon_{F} - \tilde\epsilon_{\bm p'}(\epsilon))$, we obtain
\begin{eqnarray}
  \mathrm{Im}{\Sigma}(\bm{p},\epsilon) &\approx& -\frac{\pi}{\beta}\sum_{\bm{q}}I_{\bm{p}',\bm{p}}^{(0)}(\epsilon-i\delta,\epsilon+i\delta)\delta(\epsilon + \epsilon_{F}-\tilde\epsilon_{\bm p'}(\epsilon)) \nonumber \\
  &=& -\frac{\pi}{\beta Z(\epsilon)}\lambda(\epsilon).
\end{eqnarray}

\section{Tests of convergence} \label{sec:tests-convergence}
In this Appendix, we test the convergence of our calculation.

\begin{widetext}
 \begin{minipage}{\linewidth}

\end{minipage}
\end{widetext}

\begin{figure*}[t]
\centering
 \includegraphics[width=0.85\linewidth]{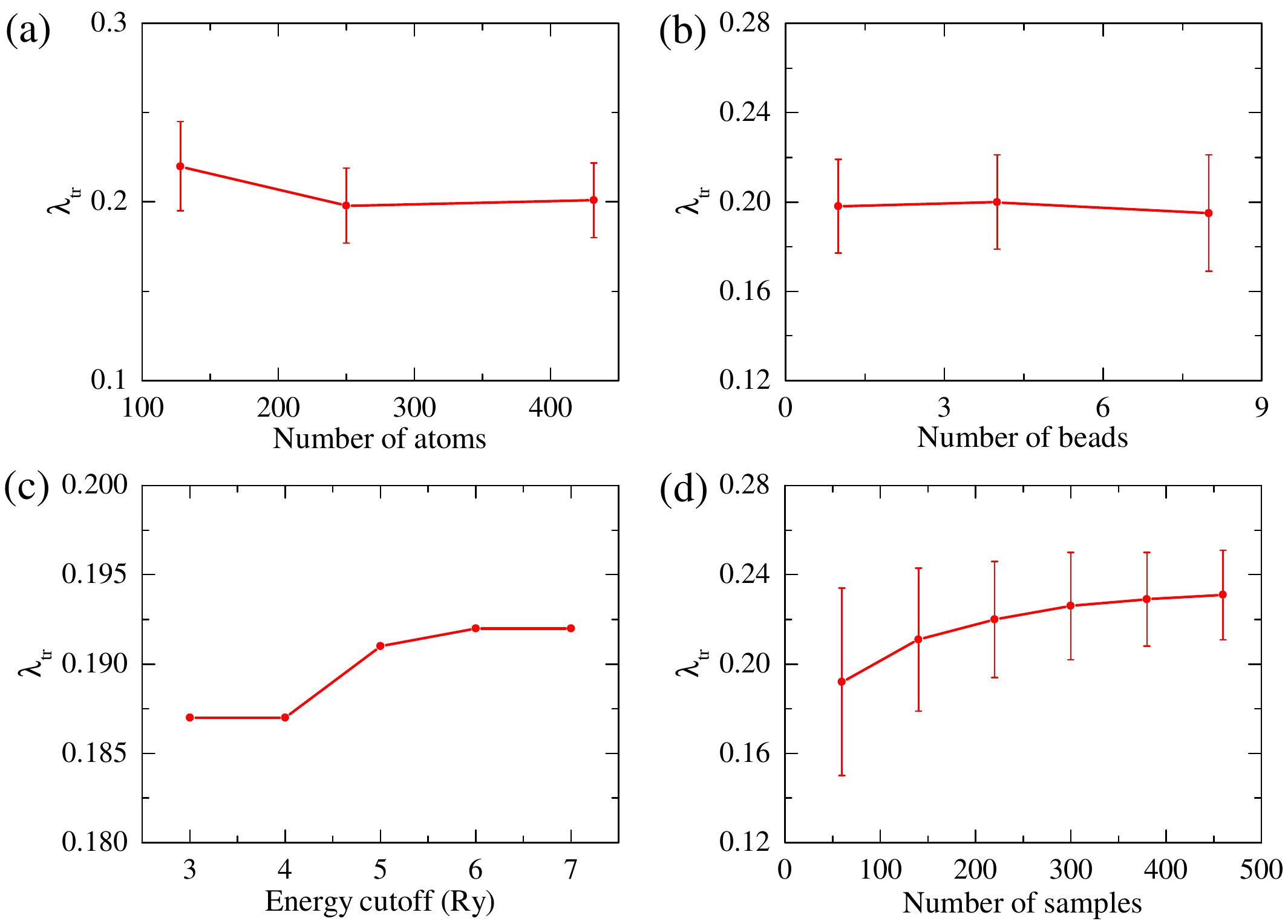}
  \caption{\label{fig-conv}The convergence test of EPC parameter $\lambda_{tr}$ based on MD and PIMD simulations at 400 K with respect to (a) the supercell size, (b) the number of beads, (c) the energy cutoff and the
  number of smaples. 250-atom supercell MD simulations, 460 samples and 6-Ry energy cutoff are chosen for the results of the main text. }
\end{figure*}

The size of the simulation supercell is the most important factor affecting the convergence. Figure~\ref{fig-conv}(a) shows the dependence of the calculated $\lambda_{\mathrm{tr}}$ at 400 K on the number of atoms in the supercell.
%
%
It can be seen that using a supercell containing 250 atoms is sufficient for the convergence of $\lambda_{\text{tr}}$. The smaller supercell makes $q$-points available for the interpolation too sparse.  The resulting uncertainty in the integrand function for determining $\lambda_{\mathrm{tr}}$ (see the inset of Fig.~\ref{fig-WGamma}) is one of the main sources of error.
%

%
%
%
The effect of quantum fluctuations is another factor being tested. We compare results from a MD simulation and PIMD simulations. Figure~\ref{fig-conv}(b) shows how the results depend on the number of the beads of the PIMD simulations (1 bead for the MD simulation).
%
We find negligible differences in the results between the MD and PIMD simulations.  This is expected since the sodium atom is heavy and the temperature (400 K) is high.
%

%
In this work, we use the plane-wave basis.
Thus the energy cutoff for these plane waves should be tested.
Fig.~\ref{fig-conv}(c) shows that $\lambda$ and $\lambda_{\text{tr}}$ are converged using an energy cutoff of 6 Ry when calculating Green's function and solving the Bethe-Salpeter equation.

%
Finally, the ensemble average of the self-energy and effective electron-electron interaction is affected by the number of sampled nuclear configurations.
Fig.~\ref{fig-conv}(d) shows that $\lambda$, $\lambda_{\text{tr}}$ and resistivity are converged by using $\sim$ 400 configurations (obtained from 10000-step MD simulations) .

\end{document}